\documentclass[superscriptaddress,twocolumn,amsmath,amssymb,aps]{revtex4-1} 
\usepackage[english]{babel} 
\usepackage[T1]{fontenc}
\usepackage{selinput}
\SelectInputMappings{%
  aacute={á},
  ntilde={ñ},
  Euro={€}
}
\usepackage{braket}
\usepackage{babel}
\usepackage{graphicx,xcolor}
\usepackage{dcolumn}
\usepackage{bm}
\usepackage{soul}
 
\usepackage{hyperref}
\hypersetup{hidelinks,
backref=true,
pagebackref=true,
hyperindex=true,
breaklinks=true,
colorlinks=true,linkcolor=black,citecolor=black,
urlcolor=blue,
bookmarks=true,
bookmarksopen=false,
pdftitle={Title},
pdfauthor={Author}}

\newcommand{\bso}{{\chi}}

\newcommand{\ipcms}{\mbox{Universit\'e de Strasbourg, CNRS, Institut de Physique et Chimie des Mat\'eriaux 
de Strasbourg,} UMR 7504, F-67000 Strasbourg, France}
\newcommand{\uba}{Departamento de F\'isica ``J. J. Giambiagi'' and IFIBA, FCEyN, Universidad de Buenos Aires, 1428 Buenos Aires, Argentina}
\newcommand{\ifimar}{Instituto de Investigaciones F\'isicas de Mar del Plata (IFIMAR), Facultad de Ciencias Exactas y Naturales, Universidad Nacional de Mar del Plata and CONICET, Funes 3350 (7600) Mar del Plata, Argentina}

\begin{document}
\title{Signatures of quantum chaos transition in short spin chains}

\author{Emiliano M. Fortes}
\affiliation{\uba}%
\author{Ignacio  Garc\'ia-Mata}
\affiliation{\ifimar} 
\author{Rodolfo  A.  Jalabert}
\affiliation{\ipcms}%
\author{Diego A. Wisniacki}
\affiliation{\uba}

\date{\today}

\begin{abstract}
The non-integrability of quantum systems, often associated with chaotic behavior, is a concept typically applied to cases with a high-dimensional Hilbert space 
Among different indicators signaling this behavior, the study of the long-time oscillations of the out-of-time-ordered correlator (OTOC) appears as a versatile tool, that can be adapted to the case of systems with a small number of degrees of freedom. Using such an approach, we consider the oscillations observed after the scrambling time in the measurement of OTOCs of local operators for an Ising spin chain on a nuclear magnetic resonance quantum simulator [J. Li,et al, Phys. Rev. X 7, 031011 (2017)]. We show that the systematic of the OTOC oscillations describes qualitatively well, in a chain with only 4 spins, the integrability-to-chaos transition inherited from the infinite chain.
\end{abstract}
\maketitle
\section{Introduction}
The Bohigas-Giannoni-Schmit (BGS) conjecture \cite{Bohigas,ullmo2016BGS} set a milestone in the study of Quantum Chaos by linking the fluctuation properties of the spectrum of a quantum system with the chaotic nature of the underlying classical dynamics. The initial numerical calculations supporting this universal connection employed restricted energy spectra. High-energy states 
were not considered because of numerical limitations, while the lowest-energy ones (typically the first 10th or 50th levels) were discarded from the statistical analysis on the premise that chaos signatures were not expected for levels close to the ground-state. Analyzing large subsets of the Hilbert space that leave aside the sector associated with the ground state has been a common practice for later work generalizing the study of level statistics to systems without classical analogue or to many-body systems \cite{ALET2018498}. Such a restriction was also adopted when considering other indicators of Quantum Chaos, like the Loschmidt echo \cite{goussev2016Loschmidt}, the Eigenstate Thermalization Hypothesis (ETH)\cite{rigol2016ETH}, and the Out-of-Time Ordered Correlator (OTOC) \cite{Swingle_Unscrambling}.

A question that naturally emerges is whether there are Quantum Chaos indicators for which it is possible to detect chaos signatures within the usually discarded low-energy sector. Or alternatively, when a small subsystem is selected from a large chaotic system, whether or not some ?memory? of the universal nature of the latter survives. In the case in which the small subsystem remains connected with the large one, the ETH provides a way to address the previous question \cite{rigol2016ETH}. 
The case of {\it isolated small systems}, where the whole spectrum is necessarily close to the ground state, constitutes the purpose of this work. 

The issue concerning the persistence (or memory) of chaos signatures in small isolated systems is not only interesting from a fundamental point of view, but also for its experimental relevance. Often, meaningful experimental results involving time-reversal protocols are obtained in systems which are considerably smaller than the ones for which the universal behavior is expected. And moreover, a reduced range of parameters could be imposed by the experimental conditions (i.e. relatively short times in order to keep quantum coherence) \cite{experimental_otoc,cappellaro2018,pastawski2019}. In the case of Ref.~\cite{experimental_otoc} a nuclear magnetic resonance quantum simulator has been developed in order to measure the OTOC of local operators for a four-site Ising spin chain. The observation that the OTOC behaves differently according to the integrability or non-integrability of the unrestricted chain stresses the importance of the previously stated question. Using a spin chain with similar parameter values than the experimental ones, we numerical show that the transition to chaos can be effectively described, despite the small number of degrees of freedom, as well as the restriction to the small times (and time-windows) attainable in the laboratory.

The transition studied in Ref. \cite{experimental_otoc} was characterized from the time behavior of the OTOC. Such a Quantum Chaos indicator can be defined \cite{Swingle_Unscrambling}, as the product of the commutator of two operators $\hat{V}$ and $\hat{W}(t)$  through
\begin{equation}
C(t) = \left\langle [\hat{W}(t),\hat{V}]^\dagger[\hat{W}(t),\hat{V}]\right\rangle \, .
\label{eq:OTOC1}
\end{equation}
The Heisenberg picture is assumed, and the angular brackets denote the average over the initial state. Thus, $C(t)$ can be interpreted as the result of the operator $\hat{V}$ probing the spread of $\hat{W}$, when the latter evolves in time. This quantity, first considered in a semiclassical theory of superconductivity \cite{1969JETP...28.1200L}, has recently been established as a measure of quantum information spreading and scrambling \cite{Bound,PhysRevA.94.040302,Riddell:2019ed,Landsman:2019ke,SwingleSlow,chen2017out,SlaglePRB2017,Luitz2017,Sahu,chen2018operator} that can be experimentally addressed \cite{experimental_otoc,garttner2017measuring,cappellaro2018,pastawski2019}.

The link of OTOC with Quantum Chaos has been developed through different steps. For systems with a large number of degrees of freedom, the initial exponential growth of $C(t)$ led to a definition of a ``quantum Lyapunov exponent'', which was shown to have a bound directly related to the system temperature \cite{Bound}. The exponential growth of the OTOC can be traced, in the case of systems with a classical analogue, to the exponential separation of trajectories in phase space \cite{hashimoto, jalabert2018semiclassical,rammensee2018many}. Even if this initial exponential growth of the OTOC has been used as a signature of Quantum Chaos
\cite{hosur2016chaos,rozenbaum2017lyapunov,chen2018operator,GarciaMata:2018kz,Hirsch2019,Sondhi2019,lakshminarayan2019out,rozenbaum2019quantum},
it has been shown that it is not a universal feature. Counterexamples appear in the case of non-integrable systems without a classical counterpart,  like in certain spin chains \cite{chen2018operator,spin2}.  

Concomitant with the limitations of focusing on the short-time behavior of the OTOC to characterize chaos, the long-time properties have been shown to exhibit the signatures of integrability and non-integrability \cite{rammensee2018many, GarciaMata:2018kz}. In particular, a new way to gauge the transition from integrability to chaos in a given system was proposed \cite{gauging}, by {\it quantifying the amplitude of the OTOC fluctuations} beyond the scrambling time. This characterization is based on the observation that the fluctuations of the OTOC after its initial short-time growth, are very large for systems were the corresponding classical dynamic is regular. And on the contrary, systems with a chaotic classical counterpart exhibit, for long times, very small amplitude oscillations that remain close to a saturation value. For systems without a classical analogue, the same systematic behavior can be established by juxtaposing the amplitude of the fluctuations to level spacing statistics and related quantum chaos indicators. The fluctuation approach has been tested for very different systems which share the common feature of a parametric transition of the dynamics from chaos to integrability, and has been shown to be successful for large times and large Hilbert space sizes \cite{gauging}. In the sequel, we apply such a method to the case of short spin chains, towards our quest for the signatures of chaos in isolated small systems.

%
%
\section{OTOC fluctuations to signal quantum chaos}
\begin{figure}
    \centering
    \includegraphics[width=0.95\linewidth]{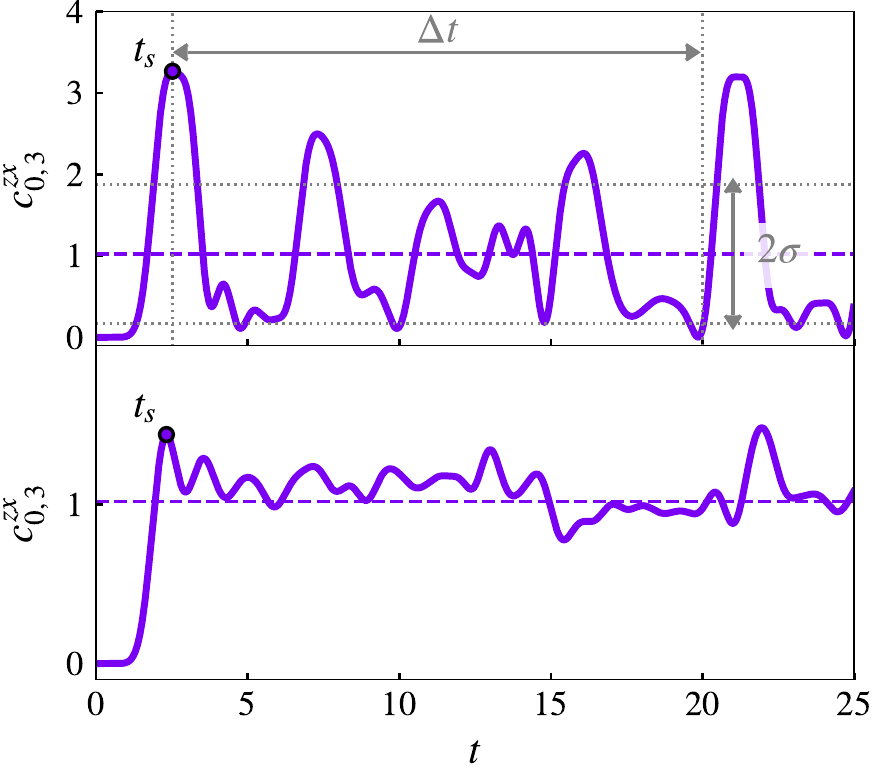}
    \caption{Two examples of the renormalized OTOC $c(t)$ as a function of time for the case of the spin chain Hamiltonian of Eq.~(\ref{hamiltonian}) with a transverse field $h_x=1$, choosing single-site Pauli operators $\hat{\sigma}_z$ at sites $0$ and $3$. The longitudinal field is $h_z=0$ (top panel) and $h_z=0.5$ (bottom panel). In the top panel, we indicate a possible time-window used to extract the value the standard deviation $\sigma$, which is the key parameter for the characterization of the integrability-chaos transition through Eq.~(\ref{eq:chaosOTOCindicator}). 
    The number of spins in the chain is $L=4$. }
    \label{fig:overview}
\end{figure}

We start our analysis by briefly describing the typical time behavior of the OTOC. For two initially commuting operators $[\hat{V},\hat{W}]=0$, the time evolving-operator spreads over the (arbitrary) operator basis and the OTOC grows. The specific law of this growth depends on the dynamics. As mentioned above, an exponential growth has been related to chaos and quantum Lyapunov exponents. This is, however not universal, e.g. for spin chains it is a power law even in the non-integrable cases \cite{Rusos,gauging}.  After the initial growth and  a transient regime, the OTOC oscillates around a constant value.
For some paradigmatic systems it has been observed \cite{gauging,rammensee2018many,GarciaMata:2018kz}  that, deep in the chaotic regime, the large time behavior is approximately constant with negligibly small fluctuations. The opposite behavior --very large fluctuations, with small number of frequencies --  is observed deep in the integrable regime \cite{Rusos,hashimoto}.  In Fig.~\ref{fig:overview} we show a graphical example of each extreme case -- integrable on the top panel and chaotic on the bottom panel for the spin chain defined by Eq.~(\ref{hamiltonian}). The two possible behaviors above described are clearly present after the initial sharp growth occurring up to the scrambling time $t_s$ (marked with a circle). For chaotic systems $t_s$ (see e.g. \cite{PhysRevA.94.040302}) is approximately the time it takes for the operator to spread over the whole basis, and for the OTOC to approach an approximately constant value. For systems with a classical counterpart it is directly related to the Ehrenfest time \cite{rozenbaum2017lyapunov,GarciaMata:2018kz}. After that
the OTOC sets around a constant average value, and the fluctuation width depends on the dynamical features. Very large amplitude oscillations can be observed in the integrable (non-chaotic) case.

Two measures were introduced in order to quantify the analogous fluctuations found in the case of long spin chains \cite{gauging}. A first one is based on the inverse of the standard deviation 
$\sigma = \sqrt{\langle {c(t)}^2\rangle - 1}$
 of the renormalized OTOC  ${c(t)}\equiv C(t)/\langle C(t)\rangle$, with the brackets standing for the time-average over a time-window $\Delta t$ (see Fig.~\ref{fig:overview}). When the fluctuations are small, $\sigma^{-1}$ is large indicating that the system more chaotic. On the other hand for integrable systems, fluctuations are large, yielding a small value of  $\sigma^{-1}$. The second method is based on the localization in Fourier space, measured by the corresponding participation ratio. Both methods were shown to yield equivalent results \cite{gauging}. For the simplicity of the presentation, we only discuss in this work the results obtained by using the first measure. And moreover, instead of considering the standard deviation $\sigma$, we compute the measure 
\begin{equation}
\label{eq:chaosOTOCindicator}
    \bso = \frac{\sigma^{-1}-\sigma^{-1}_{min}}{\sigma^{-1}_{max}-\sigma^{-1}_{min}}.
\end{equation}
We assume that the dynamics of the system can be continuously driven from regular to chaotic by changing one parameter and  $\sigma^{-1}_{min}$ ($\sigma^{-1}_{{max}}$) 
is the minimal (maximal) value obtained when sweeping over the parameter range. Therefore, $\bso \to 0$ in the integrable limit and $\bso \to 1$ in the chaotic limit. The averages and standard deviations are computed within a time window $\Delta t\equiv t_f - t_i$, with $t_i$ taken equal to or larger than the scrambling time $t_s$.
\section{OTOC based chaos measure local operators in short spin chains}
We consider an Ising spin chain described by the Hamiltonian
\begin{equation}
\label{hamiltonian}
    \hat{H}\left(J,h_x,h_z\right) = -J\sum_{i=0}^{L-2}\hat{\sigma}^z_{i}\hat{\sigma}^z_{i+1} +\sum_{i=0}^{L-1}\left(h_x \hat{\sigma}^x_i + h_z \hat{\sigma}^z_i\right),  
\end{equation}
where $L$ denotes the number of spin-1/2 sites in the chain, $\hat{\sigma}_{i}^{\mu}$ represents the spin operator at site $i=0,1,...,L-1$ with the corresponding Cartesian direction $\mu=x,y,z$. We set $\hbar=1$, such that energies are measured in units of the interaction strength $J$, and times in units of $J^{-1}$ \cite{alonso2019out}.  The parameters $h_x$ and $h_z$ are, respectively, the strength of the magnetic field in the (transverse) $x$ direction, and in the (parallel) $z$ direction. A nearest neighbor (NN) interaction has been adopted and an open boundary condition is chosen for the chain.

We now compute the OTOC and $\bso$ for the spin chain.
Selecting the Pauli spin operators $\hat{\sigma}^{\mu}_i$ for the definition of the OTOC, Eq.~(\ref{eq:OTOC1}) can be written, in the infinite temperature limit, as 
\begin{align}
    C^{\mu\nu}_{ij}(t)&=\frac{1}{2}\left\langle [\hat{\sigma}_{i}^{\mu}(t),\hat{\sigma}_{j}^{\nu}\right]^{2}\rangle\nonumber\\
&=1-\text{Re} \left\{ \text{Tr}[ \hat{\sigma}_{i}^{\mu}(t)\hat{\sigma}_{j}^{\nu}\hat{\sigma}_{i}^{\mu}(t)\hat{\sigma}_{j}^{\nu}] \right\}/D,
\label{eq:OTOC_spin}
\end{align}
where $D$ is the dimension of the Hilbert space.

\begin{figure}[h]
\includegraphics[width=.95\linewidth]{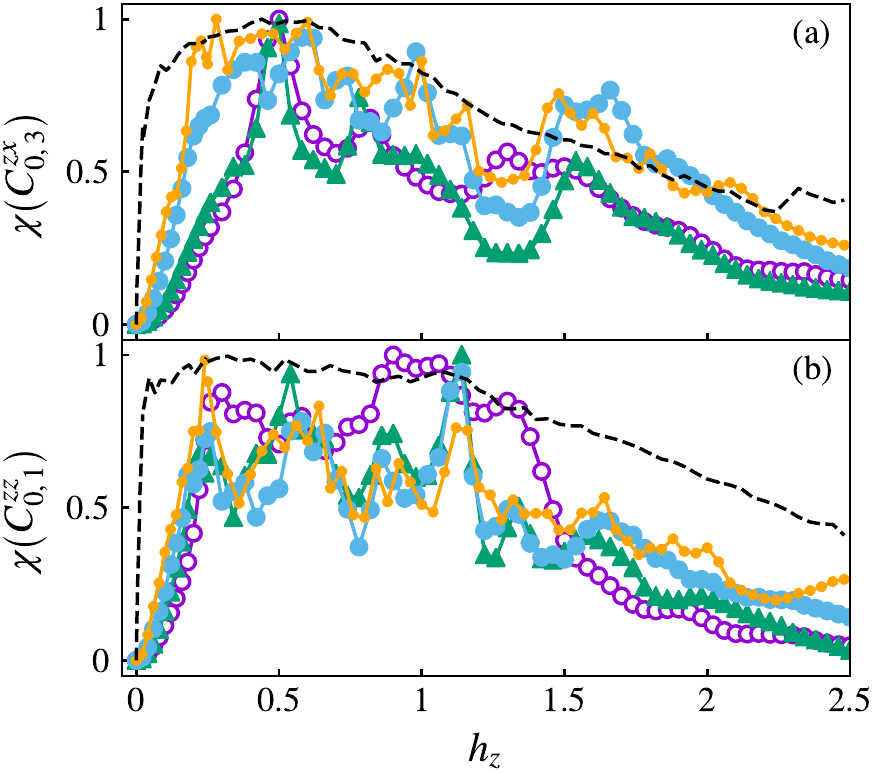}
\caption{
Parameter $\bso$ (defined in \eqref{eq:chaosOTOCindicator}) for the experimentally studied spin-chain model \eqref{hamiltonian}, using different temporal windows and two types of OTOCs: (a)$C_{0,3}^{zx}$ , and (b)$C_{0,1}^{zz}$. The averaging intervals $\Delta \tau$ are obtained, from the rescaling of 
$\Delta t = 5.5$ (empty circles), $\Delta t = 10$ (filled crosses), $\Delta t = 20$ (filled circles), and $\Delta t = 40$ (filled points). The large-time window result ($\Delta t \gg 1$) is indicated by dashed lines. \label{fig:Short_time_local_otocs_analysis}}
\end{figure}
The OTOC for the  system of Eq.~(\ref{hamiltonian}) has been successfully measured in an experiment in an NMR quantum simulator that uses the iodotrifluro-ethylene ($C_2F_3I$) molecule \cite{experimental_otoc}, where the number of  active spins is $L=4$, and the maximum operation times for the quantum evolution are quite short. The time evolution of $c(t)$ shown in Fig.~\ref{fig:overview}, corresponds to realistic experimental values. Different choices of the OTOC operators and strengths of the applied magnetic fields yield traces that, in agreement with the results of Ref. \cite{experimental_otoc}, have a different character for the integrable and the non-integrable cases. Under a fixed transverse field $h_x=1$, the difference in the long-time behavior of $c(t)$ between the regular case of $h_z=0$ (top panel) and the chaotic one of $h_z=0.5$ (bottom panel) is very clear. It is important to remark that, in our simulations, the previous difference persists well beyond times of the order of the $\Delta t$ indicated in Fig.~\ref{fig:overview} (which has been set as to correspond to the experimental case).

Using the parameter $\bso$ to characterize the transition to chaos has been proven successful in the limit of a large Hilbert space and a big $\Delta t$\cite{gauging}. As stated before, our approach is to adapt this analysis for realistic parameter values that are similar to those of Ref. \cite{experimental_otoc}. In tackling this enterprise, we first notice 
that, as $h_z$ increases, the energy spectrum spreads out. 
Therefore, to compare equivalent temporal windows  $\Delta t$
for different $h_z$ we need to rescale $\Delta t$. From the $h_z$-dependent gap 
$E_{\text{gap}}\left(h_z\right) = \text{max}\left[E\left(h_z\right)\right] - \text{min}\left[E\left(h_z\right)\right] $ ,
%
we define a scaled temporal window as  
    $\Delta \tau\left(h_z,\Delta t\right) \equiv \frac{E_{\text{gap}}\left(h_z\right) }{E_{\text{gap}}\left(0\right) } \Delta t$.
%

In Fig.~\ref{fig:Short_time_local_otocs_analysis} we show the results of $\bso$ when using different time-windows $\Delta \tau$ 
for the OTOCs $C_{0,L-1}^{zx}$ (top panel (a)) and $C_{0,1}^{zz}$ (bottom panel (b)).
The smallest $\Delta t$ is chosen on the  basis of the experimental measurements of Ref. \cite{experimental_otoc}, where it was possible to measure the local OTOCs for a time-window  $\Delta t =5.5$ (from a peak value at $t=1.5$ to a maximum time at $t=7$). 
The other time-windows have been  chosen to show 
that, by taking larger intervals, the results of $\bso$ become less noisy,
approaching  the large time-window limit ($\Delta t\gtrsim 10^3$, black-dashed lines). For all cases, it can be clearly observed that there is a steep change in $\bso$ as $h_z$ increases. The initial value of $\bso = 0$, signalling the integrable behavior for $h_z=0$ evolves in a way compatible with a regime change from integrable to chaotic. For $h_z\approx 0.5$ the parameter $\bso$ approaches a constant value.
For large enough $h_z$ ($\approx 2$) the parameter  $\bso$ decays again to zero. This behavior is easily understandable since, in the limit $h_z \gg h_x$, the Hamiltonian $\hat{H}\approx\sum_i h_z \hat{\sigma}_i^z$ is integrable. The results shown in Fig.~\ref{fig:Short_time_local_otocs_analysis} were obtained for two choices of pairs of operators (for sites $(0,1)$ and $(0,L-1)$). However, we have checked that for all the other possible choices of operators, corresponding to different sites and Cartesian directions, the results are qualitatively similar to the ones presented here (data not shown). The systematic variation of time-windows 
employed is important in order to verify that the surprising signatures of the integrable-to-chaos transition observed in the measurements and simulations of Ref. \cite{experimental_otoc} are not simply a small-size small-time effect. And thus, that the qualitatively different behavior of the two regimes persists for relatively long times, despite the small size of the system.
\begin{figure}[h]
\centering
\includegraphics[width=.95\linewidth]{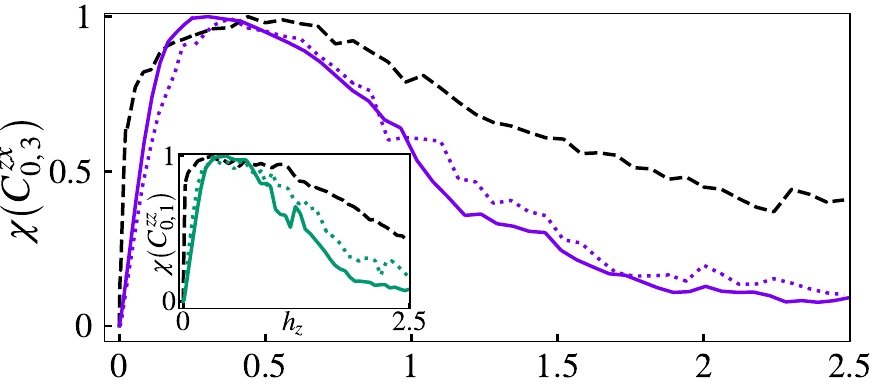}
\caption{Parameter $\bso$ for different chain lengths and two types of OTOCs: (a) $C_{0,1}^{zz}$, and (b) $C_{0,L-1}^{zx}$. The chain lengths are: $L=4$ (dashed line), $6$ (dotted line) and $8$ (solid line). The longest temporal window of Fig.~\ref{fig:Short_time_local_otocs_analysis} is used in all cases.
\label{fig:Local_otocs_long_time}}
\end{figure}


Having established that the signatures of the integrable-to-chaotic transition obtained in short chains for short times survive the consideration of longer times, we now verify that the results for $\bso$ evolve throughout a variation of the chain length in a way that is consistent with the phase transition characterizing the infinite chain. In Fig.~\ref{fig:Local_otocs_long_time} we show $\chi$ as a function of $h_z$, with a fixed $h_x=1$, for three chain lengths and different choices of the OTOC operators, using the longest time-window of the previous analysis. The dashed lines correspond to $L=4$ (also presented in  Fig.~\ref{fig:Short_time_local_otocs_analysis}), the dotted lines correspond to $L=6$ and the solid lines to $L=8$. It can be seen that the qualitative features of $\bso$ do not change upon increasing the chain length. 

\begin{figure}[h]
\includegraphics[width=.95\linewidth]{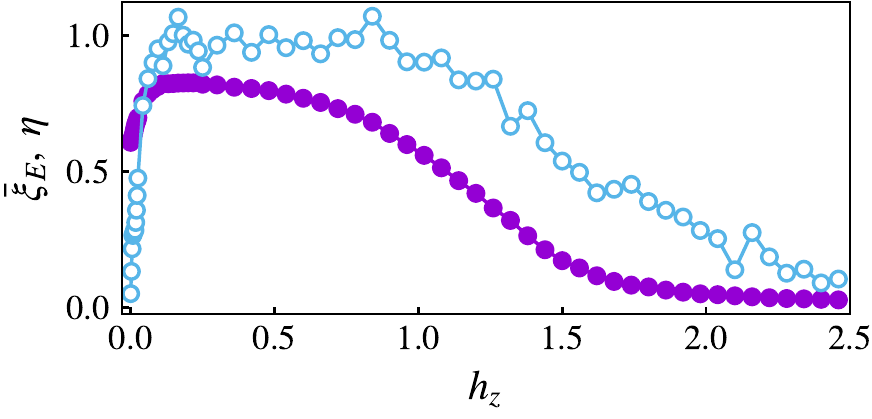}
\caption{Signature of the chaos transition from spectral measures for the spin chain model. The average PR $\bar{\xi}_{E}$ for the spin site basis  (filled circles) and the ratios measure $\eta$ (empty circles) for a chain of length $L=12$ $(D=4096)$ where we consider the even parity subspace ($D^{\text{even}}=2079$).}
\label{fig:Chaos_map}
\end{figure}

The appropriateness of the parameter $\sigma$ for detecting the chaos-to-integrability transition in very long chains was tested by benchmarking against other indicators of Quantum Chaos\cite{gauging}. In the same way, we need to test the accuracy of $\bso$ to yield the signatures of Quantum Chaos. Towards this goal, we will use two standard benchmarks. The first one, derived from the BGS conjecture and particularly useful in the case of many-body physics systems, is based on the distribution $P(\tilde{r})$, where $\tilde{r}_n={\rm min}(r_n,1/r_n)$ and  
     $r_n=(E_{n+1}-E_n)/(E_n-E_{n-1})$
 is the ratio between the two nearest-neighbor spaces of a given level $E_n$ \cite{Atas_2013}. Contrary to other spacing distributions, it does not require an energy unfolding, thus avoiding an important difficulty encountered in many-body systems, since the functional form of the level density is generally unknown. From this distribution we can define 
$\eta \equiv ({\overline{\min (1 / r, r)}-I_{P}})/({I_{\mathrm{WD}}-I_{P}})$,   
where $I_{\mathrm{WD}}\approx 0.536$ ($I_{P}\approx0.386 $) are the limiting values of $\overline{{\rm min}(r,1/r)}$ for Wigner-Dyson (Poisson) statistics. Thus, the limit $\eta\to 1$ ($\eta\to 0$) signals chaotic (regular) behavior.

As a complement of the previous spectral analysis, we consider a measure based on the eigenfunctions. We define the normalized average participation ratio (PR)
$\bar{\xi}_{E}=(D \xi_{E}^{\text {deloc }})^{-1} \sum_{i=0}^{D-1} \xi_{E_i}$ ,
where
    $\xi_{E_i}=\left(\sum_{j=0}^{D-1}\left|a_{i j}\right|^{4}\right)^{-1}$
is the PR of a single energy eigenstate $\left|E_{i}\right\rangle=\sum a_{i j}\left|\phi_{j}\right\rangle$ written in some arbitrary basis $\left\{\left|\phi_{j}\right\rangle\right\}_{j=0}^{D-1}$. 
The PR represents a  measure of localization. Large values of  $\bar{\xi}_{E}$ characterize a delocalized eigenstate which are typically associated with chaos. 
For chaotic systems $|a_{ij}|^2$ are independent random variables and
$\xi_{E}^{\text {deloc }} \approx D/3$ \cite{LS_ipr,ZELEVINSKY199685}.

In Fig.~\ref{fig:Chaos_map} we show the behavior of the parameters $\eta$ and $\bar{\xi}_{E}$ as a function of $h_z$ for the Hamiltonian of Eq.~(\ref{hamiltonian}), where we have fixed $h_x=1$. The regime change detected through the parameter $\bso$ for all values of $\Delta t$ and $L$ is equally present in the behavior of $\eta$ and $\bar{\xi}_E$ obtained for very large chains. We remark that these last two parameters rely on the statistical analysis of the spectrum. Therefore, in order to obtain valid results from  $\eta$ and $\bar{\xi}_E$, very large Hilbert spaces need to be considered (in Fig.~\ref{fig:Chaos_map} $D=4096$). By the same token, these statistical approaches cannot be applied to a simple 
$L=4$ chain 
(with a Hilbert space size $D=16$). The characterization of the OTOC fluctuations then appear as a privileged tool to address small systems close to the experimentally studied setups.

\section{OTOC chaos measure for non-local operators}
On the one hand, the OTOC measurement with local operators is an important accomplishment of Ref.~\cite{experimental_otoc}. On the other hand, for experimental NMR in solids, non-local many-particle operators are more commonly treated. Thus, the possible different behavior of the OTOC for local and non-local operators appears as an important question, also motivated by the study of settings where the scrambling in a small system depends on the interaction with a  many-particle environment \cite{niknam2020} and by the importance of non-locality in determining the short-time behavior of the echo dynamics of quantum operators \cite{pappalardi2019quantum}.

One important example of a non-local operator is the total magnetization along the direction $\mu$, 
 $\hat{\sigma}^{\mu} = \sum_{i=0}^{L-1} \hat{\sigma}_i^{\mu}$.
We will refer to the resulting correlators as mixed or global OTOCs, respectively, when they involve one or two non-local operators. And we focus on the long time-regime. Let us first consider the \emph{mixed} OTOC composed of a one-site Pauli operator $\hat{\sigma}^{\mu}_{i}$ and a total magnetization operator $\hat{\sigma}^{\nu}$. Then  
Eq.~(\ref{eq:OTOC1}) can be expanded as  
\begin{eqnarray}
    C_{i}^{\mu\nu}\left(t\right)&=&\sum_{m=0}^{L-1}C_{im}^{\mu\nu}\left(t\right)+\stackrel{L-1}{\sum_{m\neq n}}\mathcal{C}_{iimn}^{\mu\mu\nu\nu}\left(t\right)\nonumber \\    
    &\equiv& C^{\mu\nu}_{\underset{mixed}{local}}(t) +
    C^{\mu\nu}_{\underset{mixed}{non-local}}(t).
    \label{eq:MixedOTOC}
\end{eqnarray}
The first term on the right hand side corresponds to the local contribution and is composed of the OTOC defined in Eq.~(\ref{eq:OTOC_spin})
for one-site operators.
We call the second term  non-local OTOC
and  it is expressed in terms of a
four point out-of-time-oredered correlator  %
for the Pauli spin one-site operators,  which can be written (in the infinite temperature limit) as 
\begin{equation}
\begin{split}
    \mathcal{C}_{ijlm}^{\mu\nu\zeta\delta}(t)=\frac{1}{D} \text{Tr}\left[\hat{\sigma}_{i}^{\mu}\hat{\sigma}_{j}^{\nu}\hat{\sigma}_{l}^{\zeta}(t)\hat{\sigma}_{m}^{\delta}(t)\right]
    \\-\frac{1}{D}\text{Re}\left\{\text{Tr}\left[\hat{\sigma}_{i}^{\mu}\hat{\sigma}_{l}^{\zeta}(t)\hat{\sigma}_{j}^{\nu}\hat{\sigma}_{m}^{\delta}(t)\right]\right\}.
\end{split}
\label{eq:4pointOTOC}
\end{equation}

\begin{figure}[h]
\includegraphics[width=.95\linewidth]{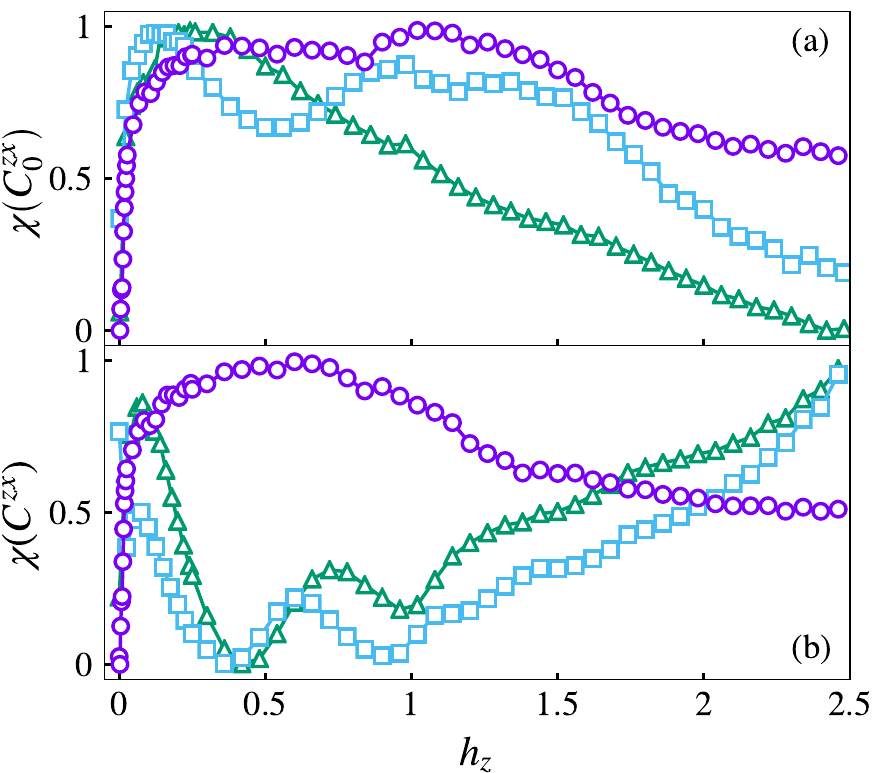}
\caption{Results of $\bso$ for the long-time temporal window in two types of mixed OTOCs. (a):  $\bso$ for the OTOC $C^{zz}_0(t)$ (light blue squares), the local contribution  (purple circles) and the the non-local contribution (green triangles). (b): The same as in panel (a) applied to the mixed OTOC $C^{zx}_0(t)$. The calculations are done for $200000$ points separated between $t_0 = 2.5$ and $t_f=5000$.}
\label{fig:mixedglobal}
\end{figure}

In fig.~\ref{fig:mixedglobal} (a) 
we show the results of $\bso$ for the mixed OTOC of operators
$\hat{V}=\hat{\sigma}^{z}_0$ and $\hat{W}=\hat{\sigma}^{x}$
with their local and non-local contribution terms from Eq.~\ref{eq:MixedOTOC}. We can see that the behavior is consistent with what is obtained in Fig.~\ref{fig:Short_time_local_otocs_analysis} for two local operators, and which is qualitatively equivalent to the behavior of the spectral chaos indicators (shown in Fig.~\ref{fig:Chaos_map}, computed using much larger Hilbert spaces).  It is clear that the general behavior is dominated by the local contribution 
$C^{zx}_{\underset{mixed}{local}}(t)$
(as was expected from the previous analysis). Although the non-local contributions introduce noise the whole $\chi$ still roughly pinpoints the transition. 

Following the same procedure, we now consider the \emph{global} OTOC for two total magnetization operators $\hat{\sigma}^{\mu}$ and $\hat{\sigma}^{\nu}$. In this case  Eq.~(\ref{eq:OTOC1})  can be expressed as 
\begin{eqnarray}
    C^{\mu\nu}(t) &=& \sum_{i,l}C_{il}^{\mu\nu}(t) + \stackrel{\forall\left(\,i\neq j\wedge l\neq m\right)}{\sum_{i}\sum_{j}\sum_{l}\sum_{m}}\mathcal{C}_{ijlm}^{\mu\nu\mu\nu}\left(t\right)\nonumber \\    
    &\equiv& C^{\mu\nu}_{local}(t) +
    C^{\mu\nu}_{non-local}(t).
    \label{eq:GlobalOTOC}
\end{eqnarray}

In fig.~\ref{fig:mixedglobal} (b) we show the  same analysis that was previously done for the mixed OTOC but applied to the global OTOC. The results of $\bso$ are presented for
$\hat{V}=\hat{\sigma}^{z}$ and $\hat{W}=\hat{\sigma}^{x}$ as well as their local and non-local contribution terms from Eq.~\ref{eq:GlobalOTOC}
Once again, a transition in accordance with the quantum chaos transition in the spectra of the system is recovered for the local contribution to the OTOC 
$C^{zx}_{local}$, but not for the non-local part
$C^{zx}_{local}$. The effect of the non-local term is more significant  than it was for the mixed OTOC, resulting in a result for $\chi$ that  does not resemble the behavior of the spectral chaos measures (Fig.~\ref{fig:Chaos_map}) at all. 
This result could impose limitations in the use of the OTOC as chaos indicator in experiments where only global magnetization measurements are available.


\section{Conclusions}
 The OTOC is a widespread quantity  used to characterize Quantum Chaos, delocalization and ergodic behavior. There are, by now, a few experiments allowing to measure the OTOC as a function of time. In this work we have focused on one experiment which simulates a spin chain. Spin chains are interesting many-body systems because they are examples of systems which are considered chaotic but might not have an initial exponential growth of the OTOC. A way to circumvent this problem is to analyze the amplitude of the OTOC fluctuations for large times (larger than the scrambling time if the system is chaotic). The method of fluctuations has previously been shown to work for very large time-windows and Hilbert space sizes, or many particles. Here we show that the method is surprisingly robust and can pinpoint the transition to chaos even in very small systems corresponding to  the regimes attainable by experiments.  We have demonstrated in this work that, for particle numbers and time-windows that are reachable by NMR experiments, the fluctuations of the OTOC are able to detect the memory of the chaotic nature characteristic of the infinite system. The resulting signatures of the chaos-to-integrable transition are qualitatively similar to the ones obtained with spectral statistics measures that require a much larger number of particles to be implemented.

Another important point that we have addressed is how the local nature of the operators considered influences the OTOC behavior. We have shown that in the OTOC for total magnetization operators, only the local part seems to have the information about the chaotic nature of the dynamics. The non-local part contributes crucially making the measure unreliable for these types of observables. This is an important results, taking into account that in most NMR experiments, the total magnetization observables are the easiest ones to measure. 
\begin{acknowledgments} 
We thank G. Alvarez for fruitful discussions.
 We received funding from CONICET (Grant No. PIP 11220150100493CO.), ANCyPT (Grant No. PICT-2016-1056), UBACyT (Grant No. 20020170100234BA), the French National Research Agency (Project ANR- 14-CE36-0007-01), a bi-national collaboration project funded by CONICET and CNRS (PICS No. 06687) and the French-Argentinian Laboratoire International Associci\'e (LIA) ``LICoQ'' (funded by CNRS).  
\end{acknowledgments}
%
%
%
%
\providecommand{\noopsort}[1]{}\providecommand{\singleletter}[1]{#1}%

\end{document}
%